\documentclass[twocolumn]{aastex631}

\usepackage{natbib}
\usepackage{multirow}
\usepackage{booktabs}
\usepackage{epsfig}
\usepackage{ulem}
\usepackage{rotating}
\usepackage{graphicx}
\usepackage{url}
\usepackage{amsmath}
\usepackage{color}
\usepackage{savesym}
\savesymbol{tablenum}
\usepackage{siunitx}
\usepackage[normalem]{ulem}
\usepackage{appendix}
\usepackage{enumerate}
\usepackage{gensymb}
\usepackage{longtable}
\usepackage{afterpage}
\usepackage{threeparttable}
\usepackage{tabularx}
\usepackage[bottom]{footmisc}

\DeclareSIUnit\parsec{pc}
\DeclareSIUnit\h{\textit{h}}

\begin{document}
    \title{The DEHVILS in the Details: Type Ia Supernova Hubble Residual Comparisons and Mass Step Analysis in the Near-Infrared}
    \author[0000-0001-8596-4746]{Erik R.~Peterson}
    \affiliation{Department of Physics, Duke University, Durham, NC 27708, USA}
    \author[0000-0002-4934-5849]{Daniel Scolnic}
    \affiliation{Department of Physics, Duke University, Durham, NC 27708, USA}
    \author[0000-0002-6230-0151]{David O. Jones}
    \affiliation{Institute for Astronomy, University of Hawai`i, 640 N. A'ohoku Pl., Hilo, HI 96720, USA}
    \author[0000-0003-3429-7845]{Aaron Do}
    \affiliation{Institute for Astronomy, University of Hawai`i at M$\bar{a}$noa, Honolulu, HI 96822, USA}
    \author[0000-0002-8012-6978]{Brodie Popovic}
    \affiliation{Department of Physics, Duke University, Durham, NC 27708, USA}
    \author[0000-0002-6124-1196]{Adam G. Riess}
    \affiliation{Space Telescope Science Institute, Baltimore, MD 21218, USA}
    \affiliation{Department of Physics and Astronomy, Johns Hopkins University, Baltimore, MD 21218, USA}
    \author[0000-0002-0800-7894]{Arianna Dwomoh}
    \affiliation{Department of Physics, Duke University, Durham, NC 27708, USA}
    \author[0000-0001-5975-290X]{Joel Johansson}
    \affiliation{Oskar Klein Centre, Department of Physics, Stockholm University, AlbaNova, SE-10691 Stockholm, Sweden}
    \author[0000-0001-5402-4647]{David Rubin}
    \affiliation{Department of Physics and Astronomy, University of Hawai`i at M$\bar{a}$noa, Honolulu, HI 96822, USA}
    \author[0000-0002-8687-0669]{Bruno O. Sánchez}
    \affiliation{Department of Physics, Duke University, Durham, NC 27708, USA}
    \author[0000-0003-4631-1149]{Benjamin J. Shappee}
    \affiliation{Institute for Astronomy, University of Hawai`i at M$\bar{a}$noa, Honolulu, HI 96822, USA}
    \author[0000-0003-2858-9657]{John L. Tonry}
    \affiliation{Institute for Astronomy, University of Hawai`i at M$\bar{a}$noa, Honolulu, HI 96822, USA}
    \author[0000-0002-9291-1981]{R. Brent Tully}
    \affiliation{Institute for Astronomy, University of Hawai`i at M$\bar{a}$noa, Honolulu, HI 96822, USA}
    \author[0000-0001-8788-1688]{Maria Vincenzi}
    \affiliation{Department of Physics, Duke University, Durham, NC 27708, USA}

\begin{abstract}
Measurements of Type Ia Supernovae (SNe Ia) in the near-infrared (NIR) have been used both as an alternate path to cosmology compared to optical measurements and as a method of constraining key systematics for the larger optical studies. 
With the DEHVILS sample, the largest published NIR sample with consistent NIR coverage of maximum light across three NIR bands ($Y$, $J$, and $H$), we check three key systematics: (i) the reduction in Hubble residual scatter as compared to the optical, (ii) the measurement of a ``mass step'' or lack thereof and its implications, and (iii) the ability to distinguish between various dust models by analyzing slopes and correlations between Hubble residuals in the NIR and optical. 
We produce SN Ia simulations of the DEHVILS sample and find that it is \textit{harder} to differentiate between various dust models than previously understood. 
Additionally, we find that fitting with the current SALT3-NIR model does not yield accurate wavelength-dependent stretch-luminosity correlations, and we propose a limited solution for this problem.
From the data, we see that (i) the standard deviation of Hubble residual values from NIR bands treated as standard candles are 0.007--0.042 mag smaller than those in the optical, (ii) the NIR mass step is not constrainable with the current sample size of 47 SNe Ia from DEHVILS, and (iii) Hubble residuals in the NIR and optical are correlated in the data. 
We test a few variations on the number and combinations of filters and data samples, and we observe that none of our findings or conclusions are significantly impacted by these modifications.

\end{abstract}
\keywords{cosmology -- type Ia supernovae -- near-infrared}

\section{Introduction}
Type Ia Supernovae (SNe Ia) are excellent indicators of our universe's expansion \citep[e.g.,][]{Freedman19,Riess22,Brout22}.
Most of the cosmological analyses with SNe have been comprised of optical light curve (LC) samples \citep[e.g.,][]{Scolnic18,Brout22,Popovic24,Vincenzi24,DES2024}; however, over the last two decades there have been a number of surveys conducted in the near-infrared (NIR) because they offer measurements with higher distance precision and can constrain systematics in the optical \citep{Meikle00,Krisciunas04,Wood-Vasey08,Phillips12,Barone-Nugent12,Avelino19,RAISIN}.
With the Dark Energy, H$_0$, and peculiar Velocities using Infrared Light from Supernovae \citep[DEHVILS;][]{Peterson23} survey data and simulations derived from these data, we analyze the reduction in scatter from optical to the NIR on the Hubble diagram; the size of the residual correlation between Hubble residuals (HRs) and host galaxy mass, ``mass step,'' in the NIR; and the relationships between HRs in the NIR and optical.

Several studies report that SNe Ia are better standard candles in the NIR, demonstrating reduced scatter on the Hubble diagram when using NIR data.
For example, \citet{Avelino19} report an RMS value of 0.117 mag from a NIR-only Hubble diagram while the corresponding optical-only values are $\sim$0.17 mag.
\citet{Peterson23} find that optical-only LCs result in 0.141 mag scatter while NIR-only LCs result in 0.102 mag scatter, and \citet{Pierel22} provide results indicating HR scatter is improved by $\sim$30\% when using NIR data rather than optical alone.

The ``mass step'' is a feature observed in the HRs where, \textit{after standardization}, SNe Ia observed in high-mass galaxies appear to be intrinsically brighter in the optical than SNe Ia found in low-mass galaxies \citep{Kelly10,Sullivan10,Lampeitl10}.
The true cause of the phenomenon is still unknown, and hypotheses for the physical explanation of the mass step vary broadly between progenitor-based and dust-based explanations \citep{Rigault13,Jones15,Jones18a,Roman18,BroutScolnic21,Johansson21,Jones23,Ye24,Grayling24,Thorp24}. 
Progenitor-based explanations attribute the mass step to either the star formation rate \citep[SFR; ][]{DAndrea11,Rigault20}, age \citep{Gupta11,Childress14,Chung23}, or metallicities \citep{Hayden13,Childress13,Rose21Mass} of the stellar populations of the host galaxies which are hypothesized to affect the composition of the progenitor or its explosion physics.
A dust-based explanation points to different models of dust properties in host galaxies \citep{BroutScolnic21,Johansson21,Popovic23}.
Since NIR light is less affected by dust \citep[e.g.,][]{Cardelli89,Fitzpatrick99,Mandel09,Mandel22}, if the mass step is due to dust, then the mass step should diminish significantly when calculated with data in the NIR \citep[e.g.,][]{Uddin20,Ponder21,Johansson21}.

Results on the existence of the NIR mass step have been inconclusive.
\citet{Uddin20} test individual filters and find a mass step ranging from $\sim$0.06--0.14 mag with errors around $\sim$0.04--0.05 mag
considering NIR filters alone.
\citet{Ponder21} find a mass step size in $H$-band of $0.13 \pm 0.04$ mag when placing the step at 10$^{10.43} M_\odot$, but they also report a mass step of $0.05 \pm 0.04$ mag with a 10$^{10} M_\odot$ step location.
\citet{Johansson21} claim to find no evidence for a NIR mass step, recovering $-0.024 \pm 0.034$ mag in $J$-band.
RAISIN \citep{RAISIN} report a NIR mass step of $0.072 \pm 0.041$ mag at 10$^{10} M_\odot$.
\citet{Thorp22} fit for mass steps using data from CSP \citep{Krisciunas17} and find NIR mass steps at various mass step locations ranging $\sim$0.02--0.08 mag with uncertainties of $\sim$0.04 mag.
Most recently, combining CSP-I and CSP-II data, \citet{Uddin23} report an $H$-band mass step of $0.014 \pm 0.057$ mag, consistent both with zero and typical step sizes in the optical.
Given that these analyses use different bandpasses covering different wavelength ranges, different LC fitters, and different analysis techniques, comparing between them to determine a definitive NIR mass step value is difficult.

Additionally, given that the effects from dust diminish in the NIR and that NIR LCs can act as better standard candles than the standardizable optical LCs \cite{Avelino19}, we can hypothesize that NIR HRs themselves may be unrelated to optical HRs.
RAISIN \citep{RAISIN} report that when comparing HRs measured using one band at a time for optical versus NIR bands, slopes are $\sim$0.78--0.86.
Results from \citet{Uddin20} indicate that HRs in the NIR are related to HRs in the optical, as HRs reported for individual bands from $B$- to $H$-band are all quite similar \citep[table A1 of][]{Uddin20}.
This hypothesis that HRs in the NIR do not relate well with HRs in the optical has not been analyzed as rigorously as the other hypotheses about scatter reduction and the mass step in the NIR, but uncovering these relationships between residuals could be a powerful indicator of the physics of SNe.

Progress has been limited with NIR studies due to the relatively small amount of quality LCs, as well as a lack of simulations for predictions.
The majority of NIR data usable for cosmological analyses comes primarily from CSP \citep{Hamuy06,Contreras10,Stritzinger11,Krisciunas17}, CfA \citep{Wood-Vasey08,Friedman15}, and DEHVILS \citep{Peterson23}, with additional data from RATIR \citep{Johansson21}, SweetSpot \citep{Weyant18}, RAISIN \citep{RAISIN}, and Hawai`i Supernova Flows \citep{Do24} among other sources. 
Here, we limit the sample in our analysis to DEHVILS data since combining samples requires extensive knowledge of each individual survey's strategies, biases, and systematics.
Further, calibration between samples must be comparable, and given that \citet{Dhawan18} find the photometric calibration for CSP and CfA results in different Hubble-flow intercepts, we opt to use solely the DEHVILS sample.
NIR SN Ia LC simulations have been constructed in \citet{Pierel22} for the SALT3-NIR model and in \citet{RAISIN} for the RAISIN sample, but cosmological analysis using NIR simulations has been limited.

In Section \ref{sec:data} we describe the data sources for the photometry, redshifts, and masses used in this analysis. In Section~\ref{sec:simulations} we detail the construction of the NIR simulations and the preparation of both the data and simulations for analysis. In Section~\ref{sec:results} we report results from the data and simulations including their impacts on Hubble diagram scatter, their resulting mass step values, and their Hubble residual comparisons.
Finally, in Sections ~\ref{sec:discussions} and~\ref{sec:conclusions}, we give our discussions and conclusions.

\section{Data}\label{sec:data}
\subsection{NIR Data}
NIR data come from DEHVILS DR1 \citep{Peterson23} where the LCs are calibrated using both 2MASS \citep{Skrutskie06} and a refinement with CALSPEC standard stars \citep{Bohlin14,Bohlin20}.
We use the final sample from \citet{Peterson23} for the bulk of our analysis here.
The sample consists of 47 SNe which have host galaxy spectroscopic redshifts and are spectroscopically-confirmed SNe Ia that make it through strict quality cuts (described in section~6 of \citet{Peterson23}). 

\subsection{Optical Data}\label{sec:opt_data}
We incorporate optical data from both ATLAS \citep{ATLAS,Smith20}\footnote{\url{https://fallingstar-data.com/forcedphot/}.} and ZTF \citep{Bellm19,Forster21} into our analysis.
ATLAS is an all-sky optical survey which covers the entire observable night sky on an approximate three-day cadence.
The telescopes used for ATLAS are equipped with broad-band optical filters $c$ for cyan (roughly a $g$+$r$ filter), and $o$ for orange (roughly an $r$+$i$ filter).
Calibration for ATLAS photometry is linked to Pan-STARRS \citep{Chambers16}, and ATLAS has a depth of $o\sim20$ mag \citep{ATLAS}.
All 47 SNe in the DEHVILS sample have a corresponding LC from ATLAS \citep{Smith20}, and peak magnitude MJDs for NIR LC fits are constrained using these ATLAS optical LCs.

ZTF is another large-area survey that covers the complete northern sky on an approximate two-day cadence.
ZTF observes with optical filters $g$, $r$, and $i$.
The survey also utilizes Pan-STARRS for calibration and reports a depth of $r\sim$ 21 mag \citep{Bellm19}.
41 of the 47 DEHVILS SNe have a corresponding LC from ZTF \citep{Forster21}.
All 41 LCs have ZTF $g$- and $r$-band data, and 20 LCs have ZTF $i$-band data that we use in this analysis.\footnote{Private communication, data to be published in Smith et al.~(in prep.).}

\subsection{Redshifts}
Spectroscopic host galaxy redshifts have been obtained for 100\% of the data sample analyzed here.
Host galaxy spectroscopic redshifts come from a variety of sources including the SuperNova Integral-Field Spectrograph \citep[SNIFS;][]{Lantz04} on the University of Hawaii 88-inch telescope, the Faint Object Camera and Spectrograph \citep[FOCAS;][]{Kashikawa02} on the Subaru telescope, and the literature\footnote{Literature values come primarily from HyperLEDA \citep[][\url{https://leda.univ-lyon1.fr/}]{HyperLEDA}.} with details provided in \citet{Peterson23}.
Peculiar velocity corrections are calculated and applied to all redshifts following \citet{Peterson22}.

\subsection{Host Galaxy Masses}\label{subsec:host_masses}

We obtain estimates for host galaxy masses, $M_*$, from optical photometry following \citet{Taylor11}.
Using optical $g$- and $i$-band photometry from both the Sloan Digital Sky Survey \citep[SDSS DR18;][]{SDSSDR18} and from Pan-STARRS1 \citep[PS1;][]{Chambers16,Flewelling20} and $K$-band photometry from 2MASS as a crosscheck, we obtain a mass estimate for all 47 SN host galaxies.
Details on how we obtain these mass estimates are provided in Appendix~\ref{appendix:obtaining_masses}.

In order to validate the use of these masses, we analyze and compare results from a second set of mass estimates calculated using \texttt{Prospector} \citep{Leja17,Johnson21}.
We use aperture-matched photometry from the ultraviolet to the NIR and fit these data with the \texttt{Prospector}-$\alpha$ model with non-parametric star formation history.
We train the SBI$++$ emulator on simulated photometry modeled after measurements from archival galaxy survey data for speed enhancements \citep{Wang23}. Our full spectral energy distribution (SED) fitting pipeline/web application will be published in McGill et al.~(in prep.).
We find that the masses from \texttt{Prospector} are largely consistent with the masses from \citet{Taylor11} (RMS of their differences is 0.18 dex), and the resulting mass step values are comparable. 
We compare the two sets of mass estimates and discuss their differences in Appendix~\ref{appendix:comparing_masses}, but we use the masses estimated following \citet{Taylor11} in the rest of this analysis.

\section{Simulations and data preparation}\label{sec:simulations}
We construct SN Ia LC simulations using \texttt{SNANA} \citep{Kessler09,Kessler19} which is a SN analysis package that can not only fit LC and cosmological parameters with data but also generate realistic simulated data.
In order to create DEHVILS simulations, we first aim to model the selection effects and follow-up efficiency for the DEHVILS survey, and we do so by deriving this information from the data itself. 
We also optimize the model selection function to match the simulated redshift distribution to the redshift distribution in the data \citep{Popovic21,Popovic24,Vincenzi23}. 
The simulated mass distribution is optimized in a similar way, and the cadence and depth characteristics are generated using the LCs of the data.
We do not simulate LC stretch ($x_1$) versus host galaxy mass correlations at this time given that $x_1$ values are too difficult to constrain when fitting NIR data alone.
Our simulations include an uncertainty on peculiar velocities, $\sigma_\textrm{PV}=250$ km/s \citep{Scolnic18,Peterson22}.
As discussed in Appendix~\ref{appendix:salt3_model_testing}, when simulating these data, we found that it was necessary to remove any $x_1$ dependence on intrinsic luminosity in order to avoid a correlation between HRs and the simulated $x_1$ values. 
This correlation resulted in NIR HRs that were much larger on average in the simulations than those derived from the data. 
This is a major limitation of the current SALT3-NIR model \citep{Kenworthy21,Pierel22} which is further discussed in Appendix~\ref{appendix:salt3_model_testing}.

\begin{figure}[!t]
    \centering
    \includegraphics[width=\columnwidth]{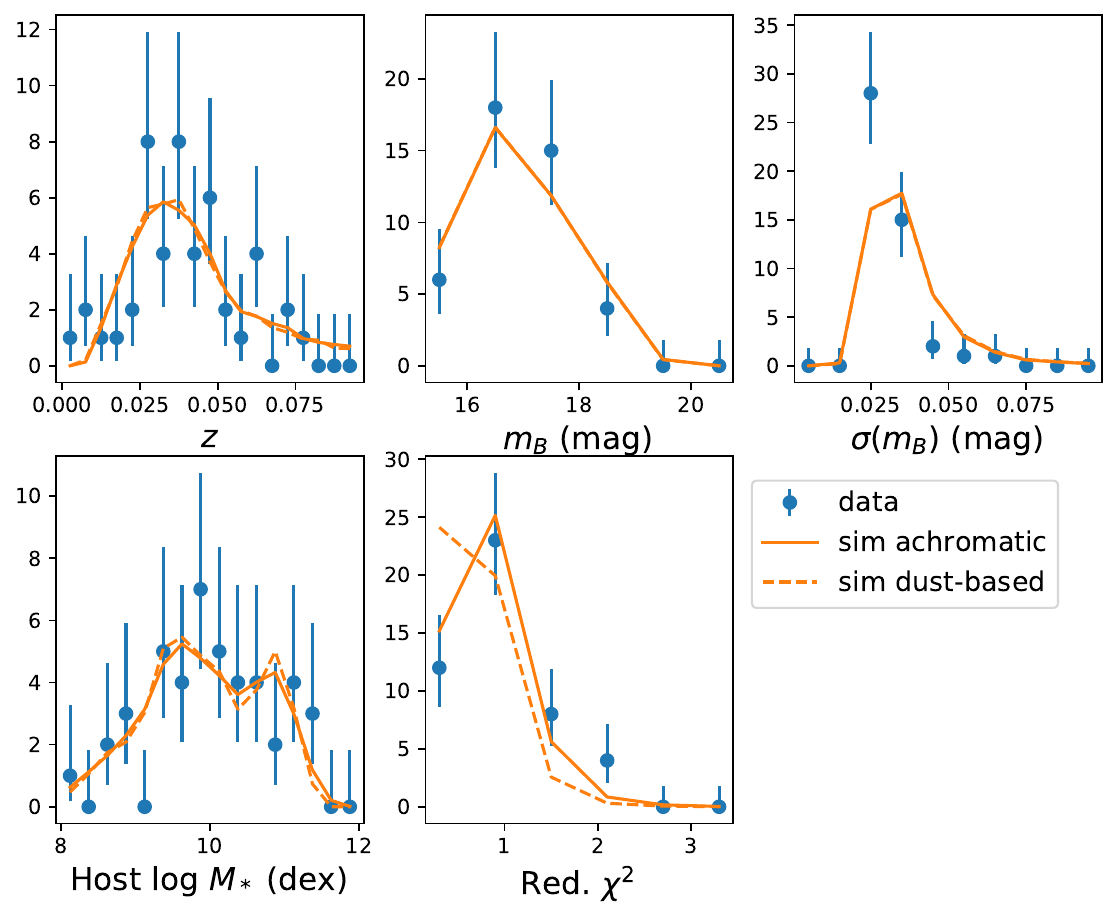}
    \caption{Binned histograms comparing various distributions from the DEHVILS data to the simulated DEHVILS data. Simulated data are normalized to the same scale as the data. Real data distributions are given in blue points with error bars derived from Poisson statistics, and simulated data are presented in orange. The distributions provided in order are redshift, apparent peak brightness $m_B$, uncertainty on apparent peak brightness $\sigma (m_B)$, host galaxy mass, and reduced $\chi^2$ of the fits.}
    \label{fig:data_vs_sims}
\end{figure}

\subsection{Intrinsic Variation Models}
In constructing these simulations, we generate two samples, each with a different model for unexplained intrinsic variations in SN brightnesses, called ``intrinsic scatter.'' We use one model with a SED variational method that models scatter as predominantly achromatic, \citet{Guy10} \citep[hereafter achromatic; e.g.,][]{Kessler13,ScolnicKessler16,Scolnic18,Bailey23}.
The achromatic model characterizes intrinsic scatter dominated by a coherent wavelength-independent scatter and is thus largely color independent (a ``gray'' scatter model).
The other model is a dust-based model framework based on \citet{BroutScolnic21} and \citet{Popovic23} (hereafter dust-based).\footnote{We follow the \citet{BroutScolnic21} framework and report a mean simulated $R_V$ of 2.1.} 
The dust-based model, in contrast, is distinctly chromatic. Color and SN Ia luminosities are based on and understood to be affected by dust, and the correlation coefficient for color, $\beta$, varies across the sample.

\begin{figure*}[!htb]
    \centering
    \includegraphics[width=\textwidth]{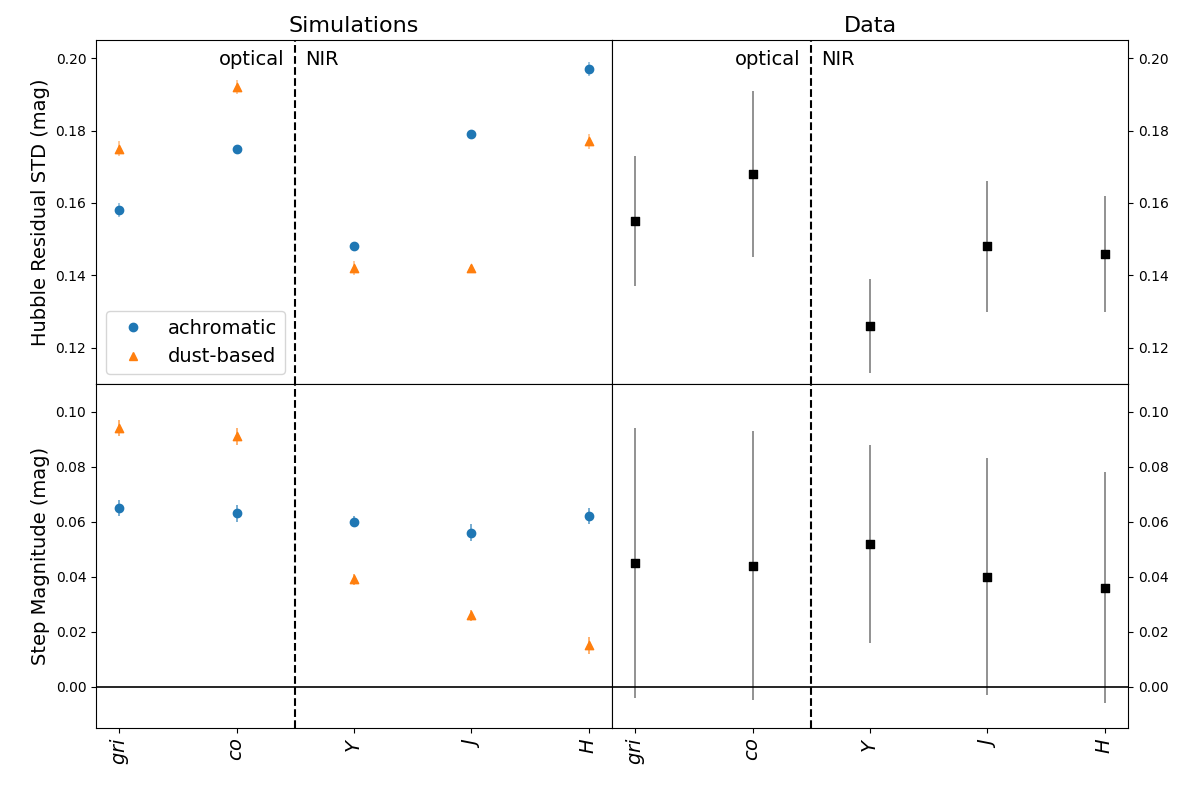}
    \caption{Visualizing Hubble residual standard deviation (STD) values and mass step values from Tables~\ref{tab:HRs} and \ref{tab:gammas} in the upper and lower panels, respectively.
    Values from simulations using the achromatic model are indicated with blue circles, and values from simulations using the dust-based model are indicated with orange triangles in the left panels. Data values are indicated with black squares in the right panels.}
    \label{fig:short_Table_vis}
\end{figure*}

\subsection{LC Fits}
Fits to both the real and simulated LCs are carried out in \texttt{SNANA} with the SALT3-NIR model \citep{Pierel22} which is an extension of the SALT3 model \citep{Kenworthy21} into the NIR.
The SALT3-NIR model parameterizes the data in order to fit the LC and a corresponding SED.
With SALT3-NIR, each LC is parameterized by a LC stretch, $x_1$, LC color, $c$, and amplitude, $x_0$, and given these parameters, a distance modulus, $\mu$, can be obtained using a modified version of the Tripp equation \citep{Tripp98},
\begin{equation}
    \mu = m_B + \alpha x_1 - \beta c - \mathcal{M},
\end{equation}

\noindent where $m_B=-2.5\log(x_0)$, the $B$-band apparent peak magnitude; $\alpha$ and $\beta$ are globally-fit nuisance parameters which relate $x_1$ and $c$ to luminosity corrections; and $\mathcal{M}$ is the SN absolute peak magnitude for a SN with $x_1,c=0,0$.

We report results and present analysis on SALT3-NIR fits to a range of filter combinations for both the real and simulated DEHVILS data in this work.
There are five sets of filters tested: three NIR-only, $Y$, $J$, and $H$,
and two optical-only, \textit{co} from ATLAS and \textit{gri} from ZTF.
A more complete set of filter combinations is analyzed in Appendix~\ref{appendix:filter_combinations}.
We fix both $x_1$ and $c$ to zero when fitting NIR-only data because $x_1$ and $c$ are largely unconstrainable with NIR data alone; we refer to this as treating the NIR SN Ia LCs as standard candles. 
We fit for both $x_1$ and $c$ for optical-only data.

\subsection{Comparing Simulations to Data}
We generate two samples of $\sim$13,000 simulated LCs that pass the same quality cuts, described in \citet{Peterson23}, as the real data: peak MJD uncertainty $< 2$ days, $E(B-V)_\textrm{MW} < 0.2$, and \texttt{SNANA}'s SN Ia LC fit probability $> 0.01$. 
In order to validate the use of the simulated DEHVILS data, we compare sample statistics from the DEHVILS data to the simulated DEHVILS data in Fig.~\ref{fig:data_vs_sims}.
For each panel, \textit{YJH}-bands are fit while fixing $x_1$ and $c$ to zero, and the simulated data are normalized to fit the data.
We compare five distributions: redshift, apparent peak brightness $m_B$, uncertainty on apparent peak brightness $\sigma (m_B)$, host galaxy mass, and reduced $\chi^2$ values.
As can be seen from Fig.~\ref{fig:data_vs_sims}, both sets of simulations (achromatic and dust-based) match the real data well; all reduced $\chi^2$ values are $<2.5$.
We are unable to reconstruct the same signal-to-noise ratio distribution with the simulated data as the real data. The data demonstrate more values with high signal-to-noise than the simulations.

Fits with optical data result in $\alpha$ and $\beta$ values of $0.162 \pm 0.043$ mag and $2.62 \pm 0.19$ mag for \textit{gri}-bands and $0.140 \pm 0.033$ mag and $2.35 \pm 0.19$ mag for \textit{co}-bands.
These $\beta$ values are smaller than typical $\beta$ values from SALT2 \citep{Guy10}, but this could be due to assumptions about uncertainties, particularly related to color.
For SALT3, both \citet{Kenworthy21} and \citet{Pierel22} note smaller $\beta$ values than from analyses using SALT2 which could be due to a constraint that requires $x_1$ and $c$ to not be correlated.
Fits to the simulations result in $\alpha$ values that are smaller and $\beta$ values that are larger than those reflected in the data. Fitted $\alpha$ and $\beta$ values from simulated optical-only data are close to 0.09 and 3.1, respectively.
We do not fit for $\alpha$ or $\beta$ with NIR data alone.

\section{Results}\label{sec:results}

We report results on each of the three hypotheses presented in the introduction with both the simulated and real samples described in this work. 
We discuss the scatter present on the Hubble diagram in Section~\ref{subsec:scatter}, our mass step analysis in Section~\ref{subsec:mass_step}, and the comparisons between HRs in the optical and NIR in Section~\ref{subsec:correlations}.

\begin{table*}[!hbt]
\caption{Hubble residual scatter values for both simulations and data.}\label{tab:HRs}
\begin{tabularx}{\textwidth}{|l| @{\extracolsep{\fill}}cccc|cc|}
\hline
 & \multicolumn{4}{c|}{Simulations} & \multicolumn{2}{c|}{Data} \\
Filters & RSD (achromatic) & STD (achromatic) & RSD (dust-based) & STD (dust-based) & RSD & STD \\
 & (mag) & (mag) & (mag) & (mag) & (mag) & (mag)\\
\hline
\textit{gri} & 0.150(2) & 0.158(2) & 0.138(2) & 0.175(2) & 0.153(32) & 0.155(19) \\
\textit{co} & 0.164(2) & 0.175(1) & 0.154(2) & 0.192(2) & 0.143(32) & 0.168(23) \\
\textit{Y} & 0.144(1) & 0.148(1) & 0.113(1) & 0.142(2) & 0.127(24) & 0.126(14) \\
\textit{J} & 0.173(2) & 0.179(1) & 0.123(1) & 0.142(1) & 0.132(23) & 0.148(18) \\
\textit{H} & 0.171(2) & 0.197(2) & 0.136(1) & 0.177(2) & 0.106(30) & 0.146(15) \\
\hline
\end{tabularx}
\end{table*}

\begin{table}[!hbt]
\caption{Mass step values for both simulations and data.}\label{tab:gammas}
\begin{tabularx}{\columnwidth}{|l|@{\extracolsep{\fill}}rr|r|}
\hline
 & \multicolumn{2}{r|}{Simulations}& Data\\
Filters & $\gamma$(achromatic) & $\gamma$(dust-based) & $\gamma$\\
 & (mag) & (mag) & (mag)\\
\hline
\textit{gri} & 0.065(3) & 0.094(3) & 0.045(49) \\
\textit{co} & 0.063(3) & 0.091(3) & 0.044(49) \\
\textit{Y} & 0.060(2) & 0.039(2) & 0.052(36) \\
\textit{J} & 0.056(3) & 0.026(2) & 0.040(43) \\
\textit{H} & 0.062(3) & 0.015(3) & 0.036(42) \\
\hline
\end{tabularx}
\textbf{Notes.~}Mass step values are calculated with a mass step location at 10$^{10} M_\odot$. The achromatic model includes an injected 0.08 mag mass step, while the dust-based model incorporates different mean $R_V$ values for low and high mass host galaxies.
\end{table}

\subsection{Hubble Residual Scatter Analysis}\label{subsec:scatter}

We provide results for the scatter present on the Hubble diagram for both the simulated and real data and find general consistency between the simulated and real data samples.
HR scatter is analyzed in two ways: (i) using the robust median absolute standard deviation (RSD) which is 1.48 multiplied by the median of the absolute value of the residual from the median residual \citep{Hoaglin00} and (ii) using the standard deviation (STD).
These statistics for both simulations and data are given in Table~\ref{tab:HRs}.
Errors on these scatter statistics are given using bootstrapping where we recalculate the given statistic, choosing residuals at random with replacement with the same sample size as the one fit, 5,000 times and report the STD of this sample as the uncertainty.

HR STD values from the simulation sample fits are plotted in the upper left panel of Fig.~\ref{fig:short_Table_vis}, and both RSD and STD values are given in Table~\ref{tab:HRs}.
Optical values are generally worse than NIR values for the dust-based model, but they are similar to if not better than the $J$- and $H$-band scatter statistics for the simulations with the achromatic model.
For both sets of simulations, the $Y$-band fits provide the best/lowest scatter values: RSD values of 0.144$\pm$0.001 mag and 0.113$\pm$0.001 mag and STD values of 0.148$\pm$0.001 mag and 0.142$\pm$0.002 mag.
Scatter statistics for $J$-band are comparable to those from $Y$-band with the dust-based model but worse for the achromatic model.

We note that distance modulus uncertainties include not only intrinsic scatter but also measurement error as well as model errors from fitting.
Intrinsic scatter (0.10-0.18 mag) is still the dominant source of uncertainty on the Hubble diagram.
When we generate alternate simulations with infinite signal-to-noise, we observe consistent HR scatter between NIR filters when using the achromatic model (STD values of $\sim$0.13--0.14), but we observe an improvement in HR scatter for redder NIR filters when using the dust-based model (STD values are 0.118, 0.096, and 0.073 mag for $Y$, $J$, and $H$, respectively).
This indicates that the larger scatter observed in $H$-band could be due to measurement noise.
Upon further investigation, we find this to be true; measurement noise in $H$-band is $>$4 times larger than in the $Y$-band for our simulations.
Additionally, measurement uncertainty in the NIR ($J$- and $H$-bands $\sim$0.04--0.07 mag) is larger than measurement uncertainty in the optical ($\sim$0.02--0.04 mag).

For the data, the HR RSD and STD values are given for the DEHVILS data sample of 47 SNe in the right portion of Table~\ref{tab:HRs}.
The STD values for the data have been plotted in the upper right panel of Fig.~\ref{fig:short_Table_vis}.
In terms of scatter, the optical values for the data are somewhat worse than the NIR values.
We observe that the best STD value comes from \textit{Y}-band fits which result in $0.126\pm0.014$ mag dispersion, similar to what we see with the simulations.
Comparing STD values in the optical to STD values in the NIR, the differences are as small as 0.007 mag (\textit{gri}-bands compared to $J$-band) and as large as 0.042 mag (\textit{co}-bands compared to $Y$-band).
In terms of RSD, the lowest amount of scatter comes from $H$-band at $0.106\pm0.030$ mag with optical-only having $\sim$$0.15\pm0.03$ mag scatter.
However, when considering the uncertainty on these scatter values ($\sim$0.02--0.03 mag), the statistics are largely consistent among NIR bands and even optical-only fits.
We further discuss these trends in HR scatter and how the underlying SALT3-NIR model factors into the simulations and data in Appendix~\ref{appendix:salt3_model_testing}.

\begin{figure*}[!htb]
    \centering
    \includegraphics[width=\textwidth]{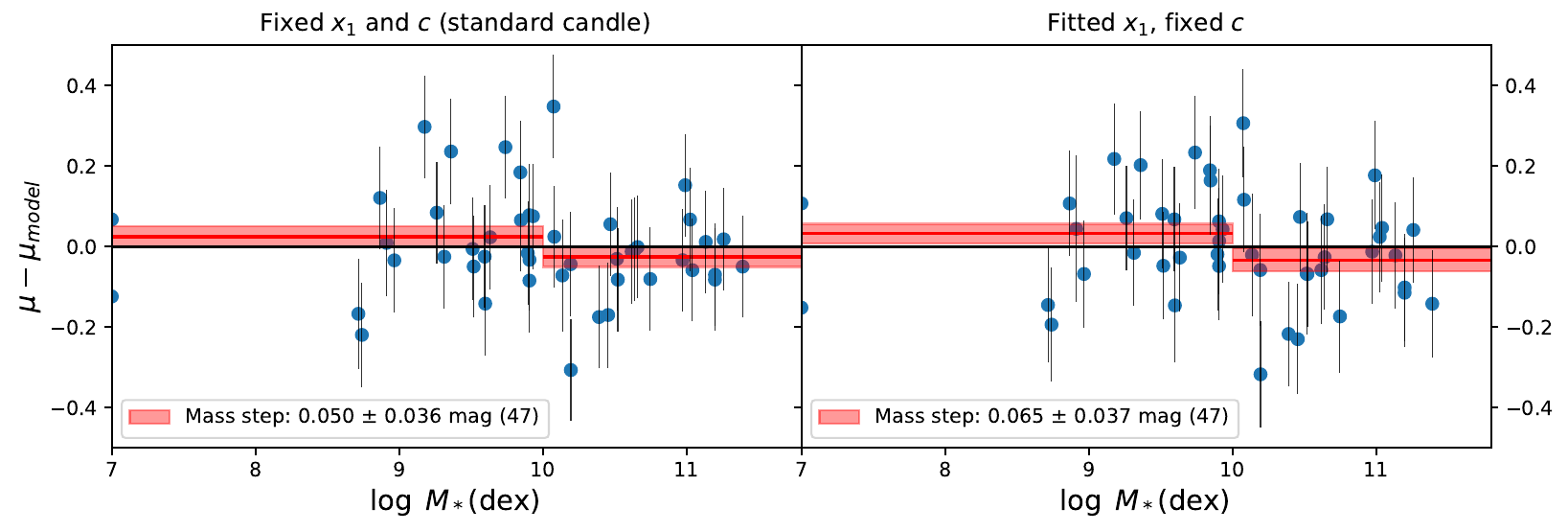}
    \caption{Mass step values from data with a mass step location at 10$^{10} M_\odot$ for two sets of \textit{YJH}-band Hubble residuals, one set fixing both $x_1$ and $c$ to zero and one set fitting for $x_1$ and fixing $c$ to zero. Both the low mass and high mass averages with standard errors are indicated in red. The number of SNe used in the calculation of the mass step is indicated in each legend.}
    \label{fig:mass_step}
\end{figure*}

\begin{figure}[!tb]
    \centering
    \includegraphics[width=\columnwidth]{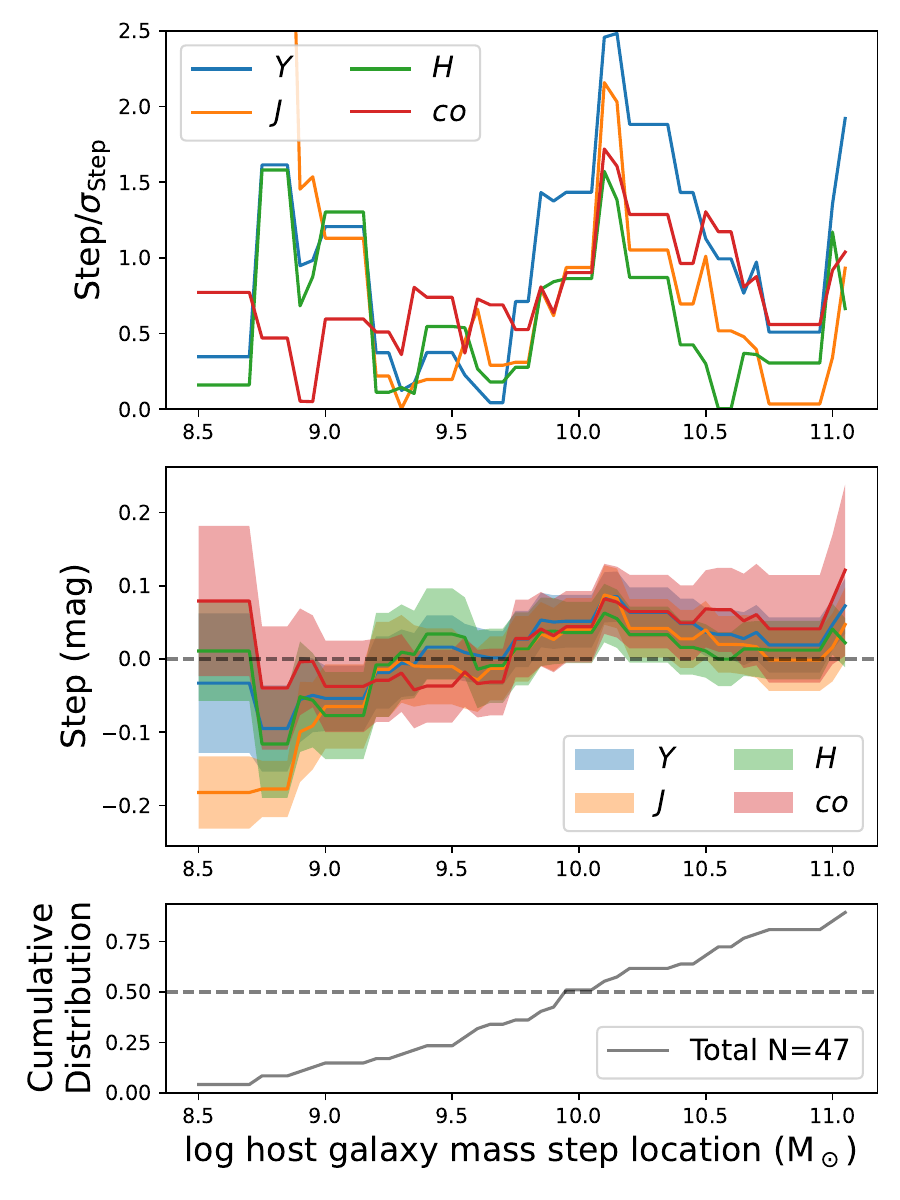}
    \caption{Mass step values and statistics from data with varying mass step locations for four sets of Hubble residuals. \textbf{Upper}: A measure of significance of the mass step, taking the magnitude of the mass step and dividing by the uncertainty. \textbf{Middle}: Each mass step with uncertainty given for each set of filters. \textbf{Lower}: A cumulative distribution function; the fraction of the sample that is below the step location at each given point.}
    \label{fig:variable_mass_step}
\end{figure}

\begin{figure*}[!htb]
    \centering
    \includegraphics[width=\textwidth]{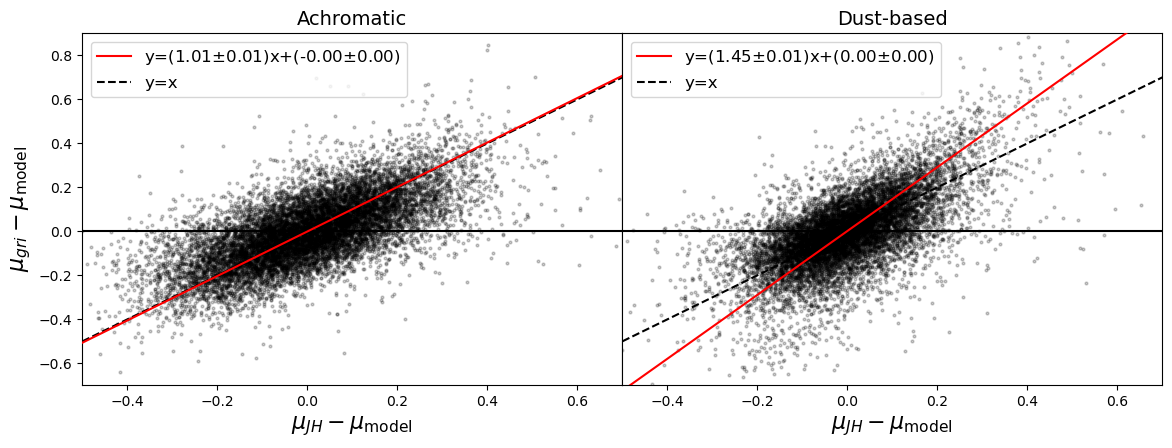}
    \caption{Hubble residual comparisons between simulated optical data (\textit{gri} with stretch and color corrections) and simulated NIR data (\textit{JH} without stretch or color corrections). Each panel is fit using orthogonal distance regression and the results are given in red. The left panel is with the achromatic model while the right panel is with the dust-based model.}
    \label{fig:HR_correlations_G10_P22}
\end{figure*}

\begin{figure}[!t]
    \centering
    \includegraphics[width=\columnwidth]{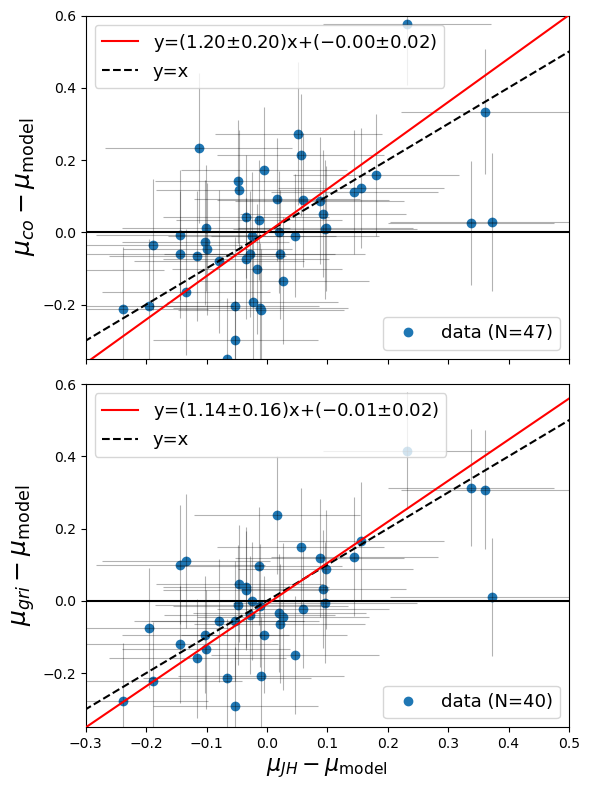}
    \caption{Similar to Fig.~\ref{fig:HR_correlations_G10_P22}, but a Hubble residual comparison between optical data from ATLAS (upper panel) and ZTF (lower panel) and NIR data from DEHVILS. Optical residuals include $x_1$ and $c$ corrections while NIR residuals do not. Each panel is fit using orthogonal distance regression and the result is provided in red.}
    \label{fig:HR_correlations_Data}
\end{figure}

\subsection{Mass Step Calculation}\label{subsec:mass_step}

We analyze the mass step, $\gamma$, by studying results from the simulations and the data each individually, comparing the simulations with the data, and varying the mass step location for the data.
We present the mass step values recovered from the simulations with a mass step location at 10$^{10} M_\odot$ in Table~\ref{tab:gammas} and the lower left panel of Fig.~\ref{fig:short_Table_vis}. 
For the achromatic model simulations, we inject a 0.08 mag mass step.
All recovered step values when using the achromatic model are relatively consistent with a mass step of $\sim$0.06 mag.
For the dust-based model simulations, where different mean $R_V$ values in low versus high mass host galaxies result in a mass step,
one of the most significant findings is that when considering the mass step for each NIR band alone we observe that none of the mass step values are zero. 
The largest mass step in the NIR is in $Y$-band, where $\gamma=0.039\pm0.002$ mag.
In $J$- and $H$-band, $\gamma=0.026\pm0.002$ and $0.015\pm0.003$ mag, respectively.
This trend in the dust-based simulation where the magnitude of the mass step decreases as a function of wavelength is expected because if the mass step itself is driven by differences in dust properties, these have less impact in the NIR \citep[e.g.,][]{Fitzpatrick99}.

In order to obtain an estimate of the difference between the simulated mass step values in the optical and in $Y$-band in terms of significance, we take a sample of 300 randomly selected SNe using bootstrapping and calculate both the optical and $Y$-band mass steps from those SNe, 5,000 times. From these distributions of mass step values, we find that, with a sample size of 300 SNe, the median $Y$-band mass step differs from the median optical $co$-band mass step by $2.0\sigma$.
Given these findings, we find that $Y$-band is still blue enough for the impact of dust to be substantial, and trying to differentiate between a dust-based color model and a gray-based color model using $Y$-band alone is challenging.
This is expected, as for e.g., a \citet{Fitzpatrick99} dust law with a total-to-selective extinction parameter, $R_V$, value of 3.1 for $V$-band, the corresponding $R_Y$ value is $\sim$1, and the corresponding $R_H$ value is $\sim$0.5.

For the data, in both the lower right panel of Fig.~\ref{fig:short_Table_vis} and Table~\ref{tab:gammas}, we observe that the uncertainties on the mass step values are large; however, the size of the mass step decreases marginally for redder wavelengths. $H$-band fits result in the smallest mass step value of 0.036$\pm$0.042 mag. 
We plot HRs as a function of host galaxy mass for two sample fits to the data in Fig.~\ref{fig:mass_step}. 
We also provide the mean HRs and standard errors for the low mass values and high mass values for a mass step location of 10$^{10} M_\odot$ in red.
When we treat the SN Ia NIR LCs as standard candles, the NIR \textit{YJH}-band mass step is $0.050 \pm 0.036$ mag in the left panel, but when we fit for $x_1$ rather than fixing it to zero, the mass step is $0.065 \pm 0.037$ mag.
This increase in the magnitude of the mass step when utilizing stretch corrections follows the findings that stretch is partially degenerate with host galaxy mass corrections \citep[e.g.,][]{RAISIN}.
We analyze the size of the mass step when varying the mass step location as well in Fig.~\ref{fig:variable_mass_step}.
In the upper panel of Fig.~\ref{fig:variable_mass_step}, the most significant positive (the typical direction of the mass step found in SN analyses) mass step is observed when the mass step is placed at $\sim$10.1 dex for all bands.
We see that when the mass step location is at or above 10.0 dex, the $Y$-band mass step divided by uncertainty is larger than that for both $J$- and $H$-bands everywhere.
Similarly, in this same range, both the optical and $Y$-band mass steps are larger at all locations than the $H$-band mass step and at every mass step location except for $\sim$10.5 dex for $J$-band.

\begin{figure*}
    \centering
    \includegraphics[width=\textwidth]{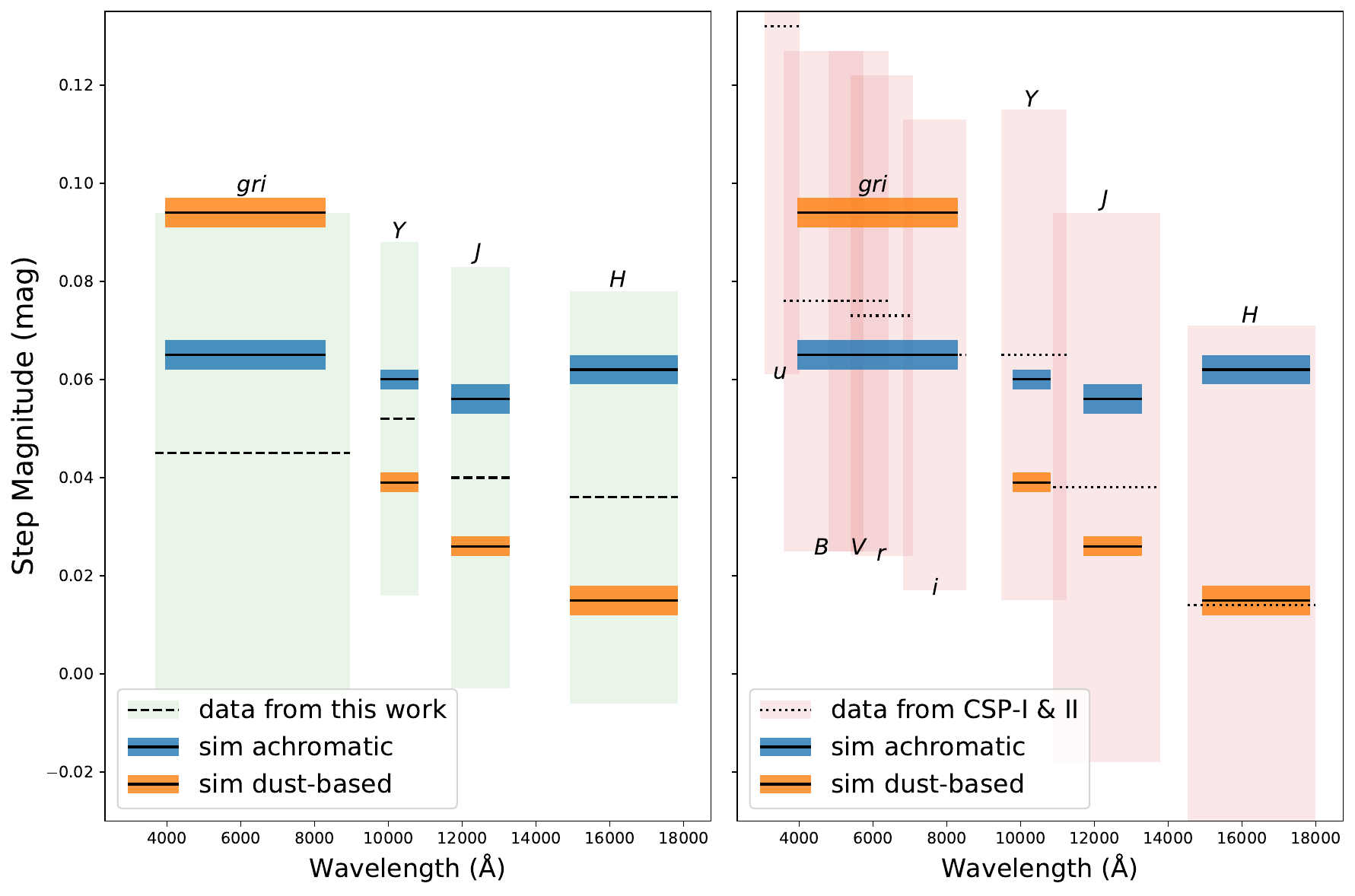}
    \caption{Comparing the calculated mass steps from simulations using both the achromatic and dust-based intrinsic scatter models to the mass steps derived from both data in this work (left panel) and data from CSP-I \& II from \citet{Uddin23} (right panel) as a function of wavelength. The mass step values from this work are calculated with a mass step location at 10$^{10} M_\odot$ while values from \citet{Uddin23} are calculated using the data sample's median mass as the mass step location (10.2--10.4 dex). The simulation values are the same in both panels, and 1$\sigma$ uncertainties are shaded for all mass step values.} 
    \label{fig:mass_step_v_wavelength}
\end{figure*}

\subsection{HR Comparisons}\label{subsec:correlations}

Given the claim that NIR SN LCs are standard candles rather than standardizable as in the optical, it can be hypothesized that
HRs in the optical are unrelated to HRs in the NIR. In order to explore this hypothesis, we compare optical simulated data HRs to HRs derived from NIR simulated data in Fig.~\ref{fig:HR_correlations_G10_P22}.
In each panel we compare \textit{gri}-band HRs to \textit{JH}-band HRs.
We use the achromatic model in the left panel and the dust-based model in the right panel.
Fitting lines using orthogonal distance regression, we find a slope of $1.01\pm0.01$ when using the achromatic model and a slope of $1.45\pm0.01$ when using the dust-based model.
For both simulations, when increasing the uncertainties on either axis by 10\%, this results in a 4\% change in the slope.

We note that we fit lines using orthogonal distance regression instead of a least-squares method, which does not consider errors on both axes.
These calculated slopes for the simulated data are consistent with the findings from the HR scatter values presented in both Table~\ref{tab:HRs} and the upper left panel of Fig.~\ref{fig:short_Table_vis}. 
The simulations with the achromatic model have comparable dispersion in the NIR and optical, while the simulations with the dust-based model demonstrate smaller dispersion in the NIR than in the optical (see Table~\ref{tab:HRs}).
These improved HRs on the x-axis of the right panel result in a larger slope relative to the left panel.

We perform a similar analysis using data in Fig.~\ref{fig:HR_correlations_Data}.
In the upper panel we compare ATLAS HRs using \textit{co}-bands to DEHVILS HRs using \textit{JH}-bands. In the lower panel we use HRs derived from ZTF \textit{gri}-band optical data. 
We fit lines using an orthogonal distance regression and find slopes of $1.20\pm0.20$ and $1.14\pm0.16$ in the upper and lower panels, respectively. 
When increasing the uncertainties on either axis by 10\%, this results in an 8\% change in the slope for the upper panel and a 5\% change in the slope for the lower panel.
Both slopes from the data are within 2$\sigma$ of and fall in between the slopes from the achromatic model simulation and the dust-based model simulation given in Fig.~\ref{fig:HR_correlations_G10_P22}.

We also compute the Pearson correlation coefficients, $\rho$, for both the simulations and data and obtain similar results between the simulations and data.
The achromatic simulated data have a Pearson $\rho$-value of 0.65 while the dust-based simulated data value is 0.64.
The Pearson $\rho$-values for the data are comparable to those from the simulations at 0.66 for \textit{gri}- versus \textit{JH}-band HRs and 0.53 for \textit{co}- versus \textit{JH}-band HRs. 
These results indicate that there is a significant correlation between HRs in the NIR and HRs in the optical.\footnote{Taking this correlation into account, when comparing mass step values between the NIR and optical, differences are small ($\sim$0.01 mag) while uncertainties are on the level of $\sim$0.04 mag.}

\section{Discussion}\label{sec:discussions}
With the dawn of the \textit{Rubin} Observatory's Legacy Survey of Space and Time \citep[LSST;][]{LSST_DC2,Sanchez22} and the \textit{Nancy Grace Roman} Space Telescope \citep{Spergel15}, it is now imperative to better understand the underlying physics of SNe Ia, distinguish between various intrinsic variation models, and uncover the true cause of the mass step.
Some work has been done to compare simulations and data with regard to the mass step as a function of wavelength, but the analysis has typically involved mass step values plotted against bandpass, allocating equal width per bandpass, rather than plotted as a function of wavelength \citep[e.g.,][]{Uddin20,Uddin23}.
We repeat the analysis from \citet{Uddin23} comparing mass step values derived from simulations with mass step values calculated from data as a function of wavelength in Fig.~\ref{fig:mass_step_v_wavelength}.
Both panels include the mass step values from Table~\ref{tab:gammas} from the simulations described in this work.
The left panel includes mass step values from DEHVILS data from Table~\ref{tab:gammas}, while the right panel presents results from CSP-I \& II data directly from table 15 of \citet{Uddin23}.
Each reported value's 1$\sigma$ uncertainty is indicated with shading, and while \citet{Uddin23} state that dust may not be responsible for the mass step, given the large uncertainties on the data, we are unable to distinguish between either intrinsic variation model.
Additionally, while \citet{Uddin23} report no variation in the mass step with respect to wavelength, by plotting the mass step values as a function of wavelength rather than allocating equal widths to all bandpasses, we can observe a general trend of a gradual decrease in the magnitude of the mass step in the data from \citet{Uddin23} in the right panel of Fig.~\ref{fig:mass_step_v_wavelength}.
We therefore conclude that we cannot rule out either the achromatic or the dust-based model with either the data from this work or the data from \citet{Uddin23}, and we are also unable to make a definitive statement on whether or not the mass step has a dependence on wavelength with either dataset.

Distinguishing between a wavelength-independent mass step, where a mass step is found to be present in the NIR, and a wavelength-dependent mass step (no NIR mass step) has been shown to be possible by \citet{Bailey23} using simulated \textit{Euclid} \citep{Euclid} and LSST data.
LSST will have rest-frame $Y$-band data; however, we find from simulations that $Y$-band data is not red enough to mitigate the effects from the mass step enough to make a definitive statement on the mass step's wavelength dependence (or independence).
With the data available currently in redder bands, constraints on the NIR mass step are too loose to make a definitive statement on the existence of a NIR mass step or lack thereof.

Given that the \textit{Nancy Grace Roman} Space Telescope will observe in the NIR (out to 23,000 \AA), we will be able to place significantly tighter constraints on the NIR mass step.
From simulations of \textit{Roman} LC data and considering the current survey strategies \citep{Hounsell18,Rose21Roman}, over the course of the complete \textit{Roman} survey, we estimate roughly 1,300 SNe Ia will be observed with rest-frame NIR data, and $\sim$300 SNe Ia will have rest-frame $J$- or $H$-band data.
With these statistics, we expect constraints on the NIR mass step to improve. Uncertainty on the NIR mass step with $J$- or $H$-band data from \textit{Roman} will reach $\lesssim$0.02 mag, and we will be better equipped to uncover the true cause of the NIR mass step.
Changes to survey strategies for the \textit{Roman} Space Telescope can improve the constraints on the NIR mass step if we are able to obtain more rest-frame NIR SN Ia data.

\section{Conclusions}\label{sec:conclusions}

In this paper, we take a close look at three hypotheses of SN Ia data in the NIR.
The first is that NIR data result in improved scatter on the Hubble diagram. 
We find from simulations that HR scatter may not decrease as expected when moving to redder wavelengths; however we do note that $Y$-band data present improved scatter compared to optical-only data.
We are unable to prove or disprove the hypothesis that HR scatter decreases for redder wavelengths with the data, given the size of the uncertainties on our scatter values.

The second hypothesis, regarding the NIR mass step, is also more difficult to distinguish than previous works have indicated. 
We find from simulations that $Y$-band is not red enough to distinguish between achromatic and dust-based explanations for the mass step.
We also see that the mass step values from simulations using both the achromatic and dust-based models are within 1$\sigma$ of the mass step values derived from the data.

Finally, we find the third hypothesis that HRs in the optical do not relate to HRs in the NIR to be unlikely.
With both simulations and data, we find significant slopes and strong correlations between NIR HRs and optical HRs.
Whether improvements in the models are necessary or the theory needs to be improved, we encourage future work be done on uncovering the reasons behind each of these results on these three important hypotheses.

By taking advantage of the NIR and the minimal impact dust has on light in this wavelength regime, we have the opportunity to more fully understand the physics of SNe Ia. However, even with larger sample sizes than current ones, teasing out differences between various scatter models using NIR data will be difficult. 
Notwithstanding, there is clear room for improvement, for example, by updating the SALT3-NIR model to better describe variations in the NIR and obtaining higher precision from the upcoming \textit{Roman} Space Telescope.
In the meantime, we encourage the community to simulate different models before concluding on the validity of a specific one.

\begin{acknowledgements}
\section*{Acknowledgements}
UKIRT is owned by the University of Hawaii (UH) and operated by the UH Institute for Astronomy.
When (some of) the data reported here were obtained, the operations were enabled through the cooperation of the East Asian Observatory.
This publication makes use of data products from the Two Micron All Sky Survey, which is a joint project of the University of Massachusetts and the Infrared Processing and Analysis Center/California Institute of Technology, funded by the National Aeronautics and Space Administration and the National Science Foundation.

D.S. is supported by Department of Energy grant DE-SC0010007, the David and Lucile Packard Foundation, the Templeton Foundation and Sloan Foundation. 
This research has made use of NASA’s Astrophysics Data System.
\section*{Software}
\texttt{SNANA} \citep{Kessler09}, {astropy} \citep{astropy:2013,astropy:2018},
{matplotlib} \citep{Hunter07},
{numpy} \citep{numpy11}, and {PIPPIN} \citep{Pippin}.
\end{acknowledgements}

\bibliographystyle{mn2e}
\bibliography{main}{}

\appendix
\section{Obtaining masses}\label{appendix:obtaining_masses}
\subsection{Masses from Optical Photometry}\label{sec:opt_masses}
In order to derive host galaxy masses from optical photometry using equation 8 from \citet{Taylor11}, 

\begin{equation}\label{eq:taylor11masses}
    \log(M_*/M_\odot) = 1.15 + 0.70(g-i)-0.4M_i,
\end{equation}

\noindent we obtain $g$- and $i$-band photometry from the Sloan Digital Sky Survey \citep[SDSS DR18;][]{SDSSDR18}\footnote{SDSS ``modelMag'' values are used as recommended by SDSS for galaxy photometry.} for the host galaxies in our sample and calculate the absolute $i$-band magnitudes, $M_i$, using distance modulus values from the redshift and a best fit cosmology.
Works using the relationship given in Eq.~\ref{eq:taylor11masses} to derive their masses often use SDSS photometry \citep[e.g.,][]{Rigault20,RAISIN}, but $\sim$50\% of our galaxies do not have $g$- and $i$-band galaxy photometry from SDSS. 
Thus, we obtain additional $g$- and $i$-band photometry from Pan-STARRS1 \citep[PS1;][]{Chambers16,Flewelling20}.\footnote{PS1 ``KronMag'' values are used given that they are defined for extended objects \citep{Kron80}.}

We compare masses obtained from SDSS photometry to masses obtained from PS1 photometry for those galaxies that have photometry in both surveys and find a relationship of $M_{*\textrm{,SDSS}} = (1.2\pm0.1) \times M_{*\textrm{,PS1}}$.
Given this relationship, for the host galaxies with PS1 photometry but no SDSS photometry, we multiply the masses calculated using PS1 photometry by 1.2 to estimate masses calculated from SDSS photometry.
In total, we obtain mass estimates for 42 out of 47 host galaxies from optical photometry.

For the five host galaxies with no photometry in either SDSS or PS1, we analyze each individually to assign them to either the low- or high-mass bins with masses of 10$^7 M_\odot$ and 10$^{12} M_\odot$, respectively.
The host galaxy for SN~2021fxy, which is NGC 5018, is an elliptical galaxy merger that falls in the high mass bin \citep{Rothberg06}.
PGC 61781, at a redshift of 0.028, is the host galaxy for SN~2021usd and has a morphological type of Sbc \citep{HyperLEDA}.\footnote{\url{http://leda.univ-lyon1.fr/}.} We deem this galaxy to be massive by visual inspection.
For SN 2020vwv, PGC 637506 has no designated morphological type, and at a redshift of 0.032, with a total $B$-magnitude of $17.4 \pm 0.5$, we deem this galaxy to be nonmassive \citep{HyperLEDA}.
The host galaxy for SN 2020kqv is PGC 9005039 and has a redshift of 0.075. We assign this galaxy to the low-mass bin.
ESO 296-17, host galaxy for SN 2020swy, is at a redshift of 0.032 with no clear designation as either massive or non-massive, so we do not assign it to a bin.
After including these four host galaxies in our mass step analysis, our total comes to 46 galaxies with masses either calculated or estimated from optical photometry.

\subsection{Masses from K-band Photometry}
We also obtain $K$-band photometry of our host galaxies from the Two Micron All Sky Survey Extended Source Catalog \citep[2MASS-XSC;][]{Jarrett00,Skrutskie06}\footnote{2MASS ``k\_m\_k20fe'' values, which are calculated using isophotal fiducial elliptical apertures, are used.} primarily as a crosscheck but also as an estimate for host galaxy masses \citep{McGaughSchombert14,Burns18} for three of the five host galaxies with no optical data available. 
Comparing $K$-band absolute magnitudes to the masses derived in Section~\ref{sec:opt_masses}, we observe a linear trend,

\begin{equation}\label{eq:K_to_Mass}
    \log(M_*/M_\odot) = (-0.43 \pm 0.03)K + (0.39 \pm 0.60),
\end{equation} 

\noindent with tight correlation, measured by a Pearson $\rho$-value of $-$0.96 at $>5\sigma$, after removing three outliers.
The three outliers come from the host galaxies of SN 2020tdy, SN 2020wgr, and SN 2021hiz.
It is likely that for SN 2020wgr and SN 2021hiz, their host galaxy masses were underestimated using Eq.~\ref{eq:taylor11masses} while the host galaxy mass for SN~2020tdy was slightly overestimated.

For the three galaxies with $K$-band photometry but without optical photometry, we use Eq.~\ref{eq:K_to_Mass} to estimate their masses.
The host galaxies for both SN 2021fxy and SN 2021usd which were designated as highly massive in the previous section are confirmed to be so with masses estimated from $K$-band photometry of 10.99 dex and 10.66 dex, respectively.
The host galaxy for SN 2020swy, which was not given a designation previously, is found to have a mass estimate of 10.20 dex here.

In the end, we construct a final photometric mass estimate compilation by taking all masses estimated from Eq.~\ref{eq:taylor11masses} except for the three outliers discussed above.
For those three outliers and the three host galaxies with $K$-band photometry but no optical photometry, we use the mass estimated from $K$-band photometry using Eq.~\ref{eq:K_to_Mass}.
Of these six host galaxies with $K$-band information incorporated into their mass estimate, the highest redshift host galaxy is that of SN~2020tdy at $z=0.04279$.
We designate the two galaxies with neither $K$-band nor optical photometry to be in the low mass bin by assigning them masses of 10$^7 M_\odot$ as discussed in Section~\ref{sec:opt_masses}.
We label this final compilation of masses $M^{47}_{*,\textrm{phot}}$; this is the sample of masses that we use in the bulk of our analysis.

\subsection{Masses from \texttt{Prospector}}
Another set of host galaxy mass estimates, which we use to further validate our masses from photometry, comes from using \texttt{Prospector} \citep{Leja17,Johnson21}.
Rather than calculating solely galaxy masses, \texttt{Prospector} is a model that aims to construct complete models for each galaxy using specifically non-parametric star formation histories and an improved statistical methodology for dealing with correlated variables.
Specifically, \texttt{Prospector} uses Flexible Stellar Population Synthesis models \citep{Conroy09} within a Bayesian hierarchical framework.
We input aperture-matched ultraviolet, optical, and NIR photometry, apply the \texttt{Prospector}-$\alpha$ model, and extract masses from the resulting galaxy models, as stated in Section~\ref{subsec:host_masses}. The full SED-fitting pipeline/web application will be provided in McGill et al.~(in prep.).
We are able to obtain masses from \texttt{Prospector} for 44 of these host galaxies, so we designate this set $M^{44}_{*,\textrm{pro}}$.

\section{Comparing masses}\label{appendix:comparing_masses}

\begin{figure*}[!htb]
    \centering
    \includegraphics[width=\textwidth]{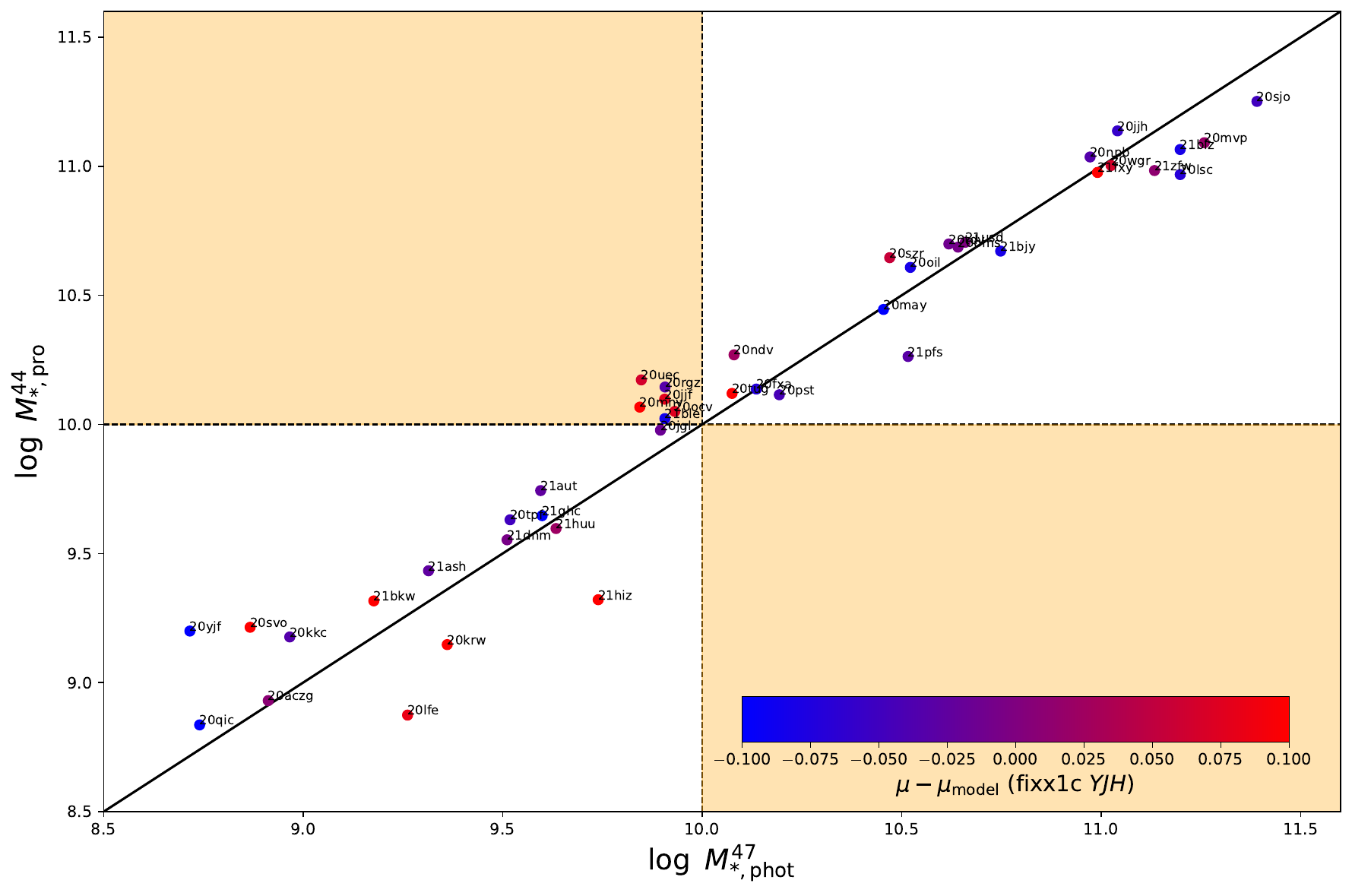}
    \caption{Comparing host galaxy mass estimates from optical and $K$-band photometry (phot) to mass estimates from \texttt{Prospector} (pro). SNe are colored by Hubble residual from NIR bands and fixing both $x_1$ and $c$ to zero. The tan regions indicate SNe which, depending on which mass estimate is used, cross the high mass/low mass threshold given a mass step location at 10$^{10} M_\odot$.}
    \label{fig:mass_v_mass}
\end{figure*}

We compare the two sets of mass estimates, $M^{44}_{*,\textrm{pro}}$ and $M^{47}_{*,\textrm{phot}}$, in Fig.~\ref{fig:mass_v_mass} where all SNe that have estimates in both sets are plotted and HRs from NIR-bands and fixed $x_1$ and $c$ are given as color.
The tan regions in the plot indicate SNe that are found in the low mass bin for one set of mass estimates and the high mass bin in the other set given a mass step location of $10^{10} M_\odot$.
Of the 43 SNe with actual mass estimates in both sets (excluding SN 2020kqv's host galaxy which is nominally given a mass of 7.0 dex), the masses are largely consistent between sets with a Pearson $\rho$-value of 0.97 at $>5\sigma$.

With the exception of the host galaxies from SN 2020rgz and SN 2021ble, all SN host galaxies that change from being high mass to low mass or low mass to high mass when switching from $M^{47}_{*,\textrm{phot}}$ to $M^{44}_{*,\textrm{pro}}$ result in a decreased mass step value.
As a result, all mass step values calculated from \texttt{Prospector} are smaller (closer to zero) in comparison to mass step values from masses calculated with photometry. 
For example, for the same fits used in Fig.~\ref{fig:mass_step}, if we use the masses from \texttt{Prospector} rather than the masses from photometry, with 44 galaxies we obtain mass step values of 0.018$\pm$0.039 when fixing $x_1$ to zero and 0.021$\pm$0.037 when fitting for $x_1$.
Masses for the complete data sample from both photometry and \texttt{Prospector} are given in Table~\ref{tab:masses}.

\begin{table}[hbt]
\centering
\caption{Host galaxy masses calculated both from photometry and \texttt{Prospector}.}\label{tab:masses}
\begin{tabularx}{0.65\columnwidth}{l@{\extracolsep{\fill}}rr}
\hline \hline
SN & $M^{47}_{*,\textrm{phot}}$ & $M^{44}_{*,\textrm{pro}}$ \\
 & (dex) & (dex)\\
\hline
2020aczg & 8.91 & 8.93 \\
2020fxa & 10.14 & 10.14 \\
2020jgl & 9.90 & 9.98 \\
2020jjf & 9.91 & 10.10 \\
2020jjh & 11.04 & 11.14 \\
2020jwl & 10.39 & -- \\
2020kkc & 8.97 & 9.18 \\
2020kqv & 7.00 & 9.48 \\
2020krw & 9.36 & 9.15 \\
2020lfe & 9.26 & 8.87 \\
2020lsc & 11.20 & 10.97 \\
2020may & 10.45 & 10.45 \\
2020mnv & 9.84 & 10.07 \\
2020mvp & 11.26 & 11.09 \\
2020ndv & 10.08 & 10.27 \\
2020npb & 10.97 & 11.04 \\
2020ocv & 9.93 & 10.05 \\
2020oil & 10.52 & 10.61 \\
2020oms & 10.64 & 10.69 \\
2020pst & 10.19 & 10.11 \\
2020qic & 8.74 & 8.84 \\
2020rgz & 9.91 & 10.14 \\
2020sjo & 11.39 & 11.25 \\
2020svo & 8.87 & 9.21 \\
2020swy & 10.20 & -- \\
2020szr & 10.47 & 10.65 \\
2020tdy & 10.62 & 10.70 \\
2020tpf & 9.52 & 9.63 \\
2020tug & 10.07 & 10.12 \\
2020uec & 9.85 & 10.17 \\
2020vwv & 7.00 & -- \\
2020wgr & 11.03 & 11.00 \\
2020yjf & 8.72 & 9.20 \\
2021ash & 9.31 & 9.43 \\
2021aut & 9.60 & 9.74 \\
2021biz & 11.20 & 11.07 \\
2021bjy & 10.75 & 10.67 \\
2021bkw & 9.18 & 9.32 \\
2021ble & 9.91 & 10.02 \\
2021dnm & 9.51 & 9.55 \\
2021fxy & 10.99 & 10.98 \\
2021ghc & 9.60 & 9.65 \\
2021hiz & 9.74 & 9.32 \\
2021huu & 9.63 & 9.60 \\
2021pfs & 10.52 & 10.26 \\
2021usd & 10.66 & 10.71 \\
2021zfw & 11.13 & 10.98 \\
\hline
\end{tabularx}
\end{table}

\section{Hubble residual scatter and the underlying SALT3-NIR model}\label{appendix:salt3_model_testing}

\begin{figure*}[!htb]
    \centering
    \includegraphics[width=0.85\textwidth]{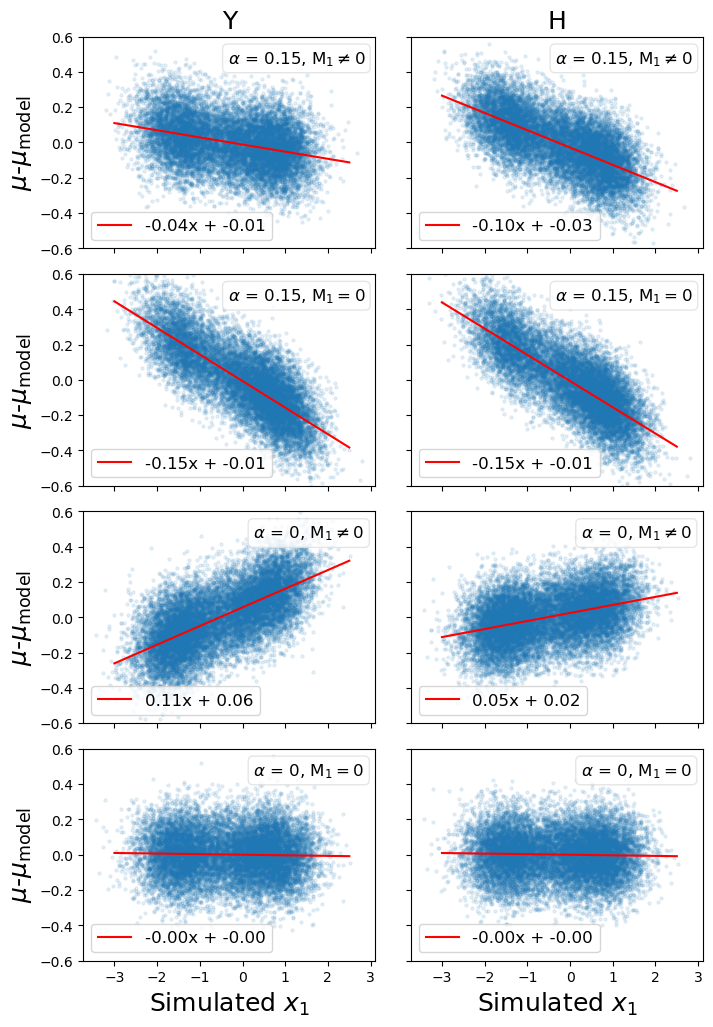}
    \caption{Hubble residuals vs.~true simulated $x_1$ values for four simulation variations in both $Y$-band and $H$-band. When simulating these samples we toggle two components, $\alpha$ and the $M_1$ model surface, on and off in each row and observe the effect on $Y$-band fits in the left panels and $H$-band fits in the right panels fixing $x_1$ and $c$ to zero. We fit the SN LCs the same way in all panels. Lines in red are fit using orthogonal distance regression in all panels and the results are provided in the legends.}
    \label{fig:splitalpha_motivation}
\end{figure*}

An important finding of \citet{Burns18} is that the stretch-luminosity slope decreases significantly for redder wavelengths. This can be seen in tables~1 and 3 of \citet{Burns18} where $P^1$, the first order coefficient to the luminosity correction due to LC stretch, goes from $\sim$0.9 in the optical to $\sim$0.1--0.3 in the NIR.
This decrease in the stretch-luminosity slope is also apparent in fig.~2 of \citet{Burns18} where the (color-corrected) $H$-band luminosity as a function of $s_{BV}$ is largely flat for $s_{BV}$ values near 1 (0.8 $<s_{BV}<$ 1.3) while for the same function but with $B$-band, the slope is nonzero. The \texttt{BayeSN} \citep{Mandel22,Ward23} LC fitting model has a similar approach where the impact of the $\theta_1$ parameter, their stretch or ``shape'' parameter, directly propagates to shifts in magnitude, dependent on wavelength and phase.

With the SALT model framework, there is a different conceptual approach where the model surface stretch ($M_1$) component is centered around zero and not a magnitude, unlike for other LC fitting models such as SNooPy \citep{Burns11,Burns14} or \texttt{BayeSN} \citep{Mandel22,Ward23}. 
Independently, the $\alpha$ parameter achromatically offsets the luminosity of the SN.
This poses a challenge for how to model a wavelength-dependent stretch-luminosity slope within the SALT model framework. 
When we fit our data in the NIR while fixing both $x_1$ and $c$ to zero we observe similar, if not better, HR scatter results than in the optical.
In contrast, when we originally fit our simulated SNe, which were simulated using the SALT3-NIR model and an $\alpha=0.15$, we observed a stark increase in HR scatter for redder wavelengths (HR STD values for $H$-band reached $\sim$0.24 mag) for both the achromatic and dust-based models.
Therefore, there is an inconsistency between the data and the simulations.

To explore this issue, we simulate a number of samples with the SALT3-NIR model turning off and on either the $M_1$ component or the $\alpha$ component in the simulations in Fig.~\ref{fig:splitalpha_motivation}.
Each panel essentially details the relative importance of stretch on intrinsic luminosity; the larger the slope, the more stretch affects the overall luminosity.
When we simulate our SNe in the NIR leaving $\alpha=0.15$ and the SALT3-NIR model unchanged and then fit our SN LCs while fixing $x_1$ and $c$ to zero, we observe HRs with large dispersion and a significant slope between those HRs and the true $x_1$ values in the first row of Fig.~\ref{fig:splitalpha_motivation}.
Specifically, when we fit a line to these HRs using orthogonal distance regression, we see that the slope of the $H$-band residuals is more than double that of the $Y$-band residuals.
The slope that we observe in $H$-band is unlike what we have seen with real data or in other LC models where the stretch-luminosity slope is near zero.

In the second row of Fig.~\ref{fig:splitalpha_motivation}, we zeroed out the SALT3-NIR model surface for $M_1$ in the NIR and retrieved a slope of magnitude 0.15, as expected. In this case, \texttt{SNANA} simulates a magnitude offset given the true $\alpha$ and $x_1$ values which are then not corrected for in the fitting stage.

In the third row, we forced $\alpha=0$ and left the $M_1$ model surface unmodified, and we are able to observe the impact that the $M_1$ model surface has on the luminosity as a function of $x_1$. For $Y$-band, the impact is large, while for $H$-band, the impact is less pronounced. We might expect that the slope for this same test in the optical would be essentially flat because in the optical, SN LCs are offset in luminosity fully by $\alpha$.
For $H$-band however, where the stretch-luminosity slope has been found to be small, this panel demonstrates that the $M_1$ model surface is not as it should be. 
The impact from this model surface should instead almost exactly counteract the impact from $\alpha$ because the intrinsic luminosity of an $H$-band LC is not as dominated by the LC stretch as it is in the optical.
Either this model surface should be modified to represent the data better, or we need to explore the option of a wavelength dependent $\alpha$ in the SALT framework.

Finally, in the fourth row, when we turn off both $\alpha$ and the $M_1$ model surface and then fit while fixing $x_1$ and $c$ to zero we observe no correlation between the simulated $x_1$ value and HRs, and this is the framework we use for our NIR simulated samples in the analysis of this work. We observe that fits to these simulations result in scatter values that are much more consistent with the data in Fig.~\ref{fig:short_Table_vis}.
We recognize that this is only a temporary solution to this problem of simulations in the NIR. We encourage additional investigation into what could be causing an inaccurate representation of the $M_1$ model surface in the SALT3-NIR model (more spectra in $H$-band might be necessary in order to produce an accurate $M_1$ model surface) or the possible inclusion of a wavelength-dependent $\alpha$ into SN Ia analyses using SALT.

Even further, we find that \texttt{SNANA} takes a given intrinsic luminosity from the user-defined distribution but then not only (i) injects a luminosity offset given the sample's true $\alpha$ parameter and each SN's true $x_1$ value but also (ii) injects a luminosity offset given the integration of the SALT model surface for $M_1$ for a given bandpass multiplied by the SN's true $x_1$ value.
This is a problem, for example, when fitting for $H$-band with simulations, where the $x_1$ parameter is ill-constrained (see fig.~7 of \citet{Pierel22}), and so accounting for this luminosity offset injected into the simulated intrinsic luminosity is near impossible.
Given the current framework for NIR simulations with \texttt{SNANA} and the SALT framework, it is assumed that the $x_1$ parameter can be well-constrained, but this is difficult with NIR data alone.

To avoid this issue for this work, we simulate the optical and NIR bands with two separate $\alpha$ values --- in the optical at $0.15$ and in the NIR at $0$. 
This requires two separate simulated samples that must be combined afterwards, and should be considered a temporary solution.

\section{Expanded filter and fitting parameter variations}\label{appendix:filter_combinations}

\begin{figure*}[!htb]
    \centering
    \includegraphics[width=\textwidth]{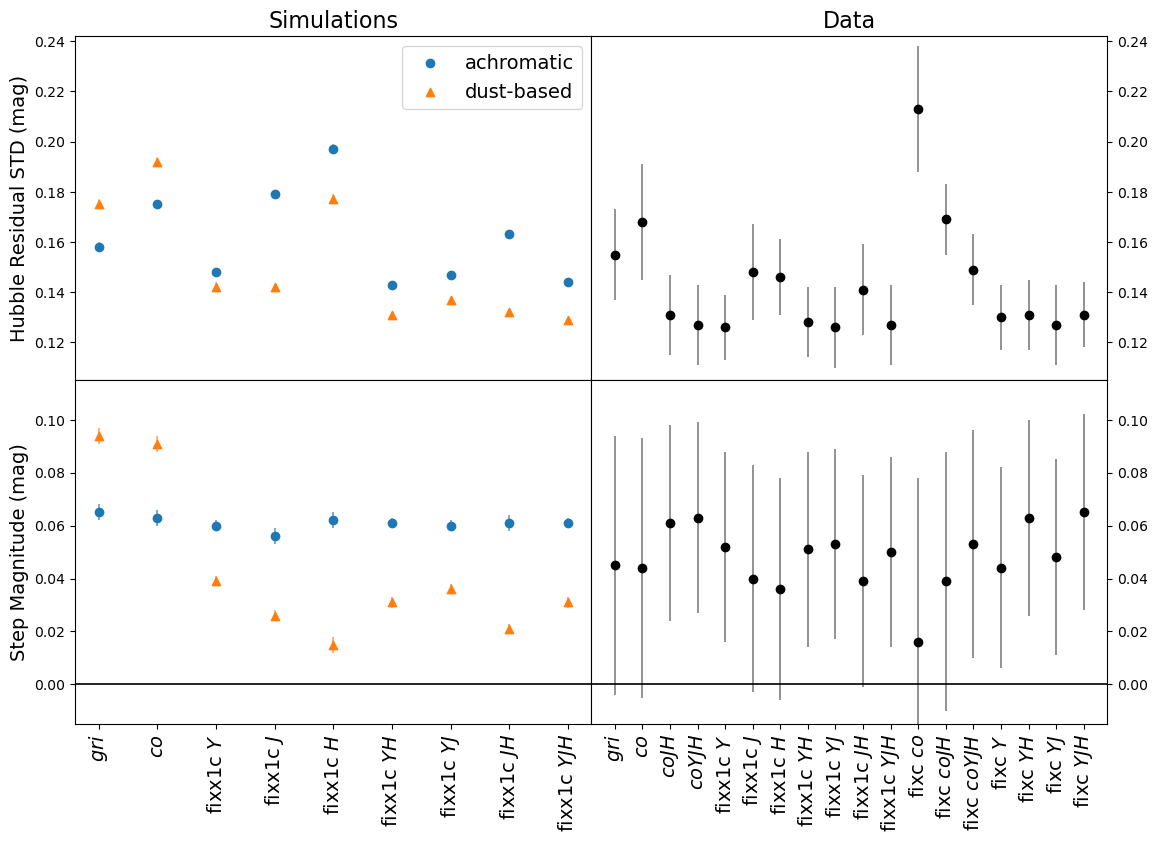}
    \caption{Visualizing Hubble residual STD values (upper panels) and mass step values (lower panels) from simulations in Tables~\ref{tab:long_G10} and \ref{tab:long_P22} (left panels) and data in Table~\ref{tab:long_Data} (right panels).
    Nomenclature in the x-axis labels goes as follows: filters fit are given along with a descriptor such as ``fixx1c'' which corresponds to fixed $x_1$ and $c$, ``fixc'' which corresponds to fixed $c$ and fitted $x_1$, or no descriptor which corresponds to fitted $x_1$ and $c$.
    This is an expansion of Fig.~\ref{fig:short_Table_vis}.}
    \label{fig:long_Table_vis}
\end{figure*}

Here, we aim to report results and present observations on a large range of filter combinations as an extension of Fig.~\ref{fig:short_Table_vis} and Tables~\ref{tab:HRs} and~\ref{tab:gammas}.
We provide results from a number of extra simulated samples in Tables~\ref{tab:long_G10} and~\ref{tab:long_P22} and the left panels of Fig.~\ref{fig:long_Table_vis} for both the achromatic and dust-based intrinsic scatter models.
For the data, we fit 11 distinct sets of filters in total.
We include seven with NIR filters only; $Y$, $J$, $H$, \textit{YJ}, \textit{YH}, \textit{JH}, \textit{YJH}.
We analyze two sets of optical-only data, \textit{gri} from ZTF and \textit{co} from ATLAS.
Combining optical with NIR data, we present results from two different combinations, \textit{coJH} and \textit{coYJH}.
In addition to this variety of filter combinations, we test a variety of fitting parameters as well, specifically we test 3 sets of fitting criteria, fixing both $x_1$ and $c$, fixing $c$ while fitting $x_1$, and fitting both $x_1$ and $c$. 
We provide results from these fits to the data in Table~\ref{tab:long_Data} and the right panels of Fig.~\ref{fig:long_Table_vis}.
None of the results from either the tables or figure deviate significantly from the findings in this paper.

\begin{table*}[!hbt]
\caption{Simulated data SALT3-NIR fit parameters, mass step values, and Hubble residual scatter using an achromatic model.}
\begin{tabularx}{\textwidth}{l @{\extracolsep{\fill}} ccccccccc}
\hline \hline
Filters & $x_1$ & $c$ & $\alpha$ & $\beta$ & $\gamma$ & N fit & RSD & STD \\
 & & & & & (mag) & & (mag) & (mag) \\
\hline
\textit{gri} & fitted & fitted & 0.093 $\pm$ 0.001 & 3.34 $\pm$ 0.02 & 0.065 $\pm$ 0.003 & 12975 & 0.150 $\pm$ 0.002 & 0.158 $\pm$ 0.002 \\
\textit{co} & fitted & fitted & 0.100 $\pm$ 0.001 & 3.14 $\pm$ 0.02 & 0.063 $\pm$ 0.003 & 13117 & 0.164 $\pm$ 0.002 & 0.175 $\pm$ 0.001 \\
\textit{Y} & fixed & fixed & 0.000 & 0.00 & 0.060 $\pm$ 0.002 & 13623 & 0.144 $\pm$ 0.001 & 0.148 $\pm$ 0.001 \\
\textit{J} & fixed & fixed & 0.000 & 0.00 & 0.056 $\pm$ 0.003 & 13617 & 0.173 $\pm$ 0.002 & 0.179 $\pm$ 0.001 \\
\textit{H} & fixed & fixed & 0.000 & 0.00 & 0.062 $\pm$ 0.003 & 13503 & 0.171 $\pm$ 0.002 & 0.197 $\pm$ 0.002 \\
\textit{YH} & fixed & fixed & 0.000 & 0.00 & 0.061 $\pm$ 0.002 & 13624 & 0.142 $\pm$ 0.001 & 0.143 $\pm$ 0.001 \\
\textit{YJ} & fixed & fixed & 0.000 & 0.00 & 0.060 $\pm$ 0.002 & 13623 & 0.146 $\pm$ 0.001 & 0.147 $\pm$ 0.001 \\
\textit{JH} & fixed & fixed & 0.000 & 0.00 & 0.061 $\pm$ 0.003 & 13624 & 0.158 $\pm$ 0.001 & 0.163 $\pm$ 0.001 \\
\textit{YJH} & fixed & fixed & 0.000 & 0.00 & 0.061 $\pm$ 0.002 & 13624 & 0.142 $\pm$ 0.001 & 0.144 $\pm$ 0.001 \\
\hline
\label{tab:long_G10}
\end{tabularx}
\end{table*}

\begin{table*}[!ht]
\caption{Simulated data SALT3-NIR fit parameters, mass step values, and Hubble residual scatter using a dust-based model.}
\begin{tabularx}{\textwidth}{l @{\extracolsep{\fill}} ccccccccc}
\hline \hline
Filters & $x_1$ & $c$ & $\alpha$ & $\beta$ & $\gamma$ & N fit & RSD & STD \\
 & & & & & (mag) & & (mag) & (mag) \\
\hline
\textit{gri} & fitted & fitted & 0.090 $\pm$ 0.001 & 3.07 $\pm$ 0.01 & 0.094 $\pm$ 0.003 & 11672 & 0.138 $\pm$ 0.002 & 0.175 $\pm$ 0.002 \\
\textit{co} & fitted & fitted & 0.089 $\pm$ 0.001 & 2.94 $\pm$ 0.01 & 0.091 $\pm$ 0.003 & 11751 & 0.154 $\pm$ 0.002 & 0.192 $\pm$ 0.002 \\
\textit{Y} & fixed & fixed & 0.000 & 0.00 & 0.039 $\pm$ 0.002 & 12980 & 0.113 $\pm$ 0.001 & 0.142 $\pm$ 0.002 \\
\textit{J} & fixed & fixed & 0.000 & 0.00 & 0.026 $\pm$ 0.002 & 12976 & 0.123 $\pm$ 0.001 & 0.142 $\pm$ 0.001 \\
\textit{H} & fixed & fixed & 0.000 & 0.00 & 0.015 $\pm$ 0.003 & 12863 & 0.136 $\pm$ 0.001 & 0.177 $\pm$ 0.002 \\
\textit{YH} & fixed & fixed & 0.000 & 0.00 & 0.031 $\pm$ 0.002 & 12980 & 0.109 $\pm$ 0.001 & 0.131 $\pm$ 0.001 \\
\textit{YJ} & fixed & fixed & 0.000 & 0.00 & 0.036 $\pm$ 0.002 & 12978 & 0.110 $\pm$ 0.001 & 0.137 $\pm$ 0.001 \\
\textit{JH} & fixed & fixed & 0.000 & 0.00 & 0.021 $\pm$ 0.002 & 12980 & 0.116 $\pm$ 0.001 & 0.132 $\pm$ 0.001 \\
\textit{YJH} & fixed & fixed & 0.000 & 0.00 & 0.031 $\pm$ 0.002 & 12979 & 0.107 $\pm$ 0.001 & 0.129 $\pm$ 0.001 \\
\hline
\label{tab:long_P22}
\end{tabularx}
\end{table*}

\begin{table*}[!hbt]
\caption{Data SALT3-NIR fit parameters, mass step values, and Hubble residual scatter.}
\begin{tabularx}{\textwidth}{l @{\extracolsep{\fill}} cccccccccc}
\hline \hline
Filters & $x_1$ & $c$ & $\alpha$ & $\beta$ & $\gamma$ & N fit & RSD & STD \\
 & & & & (mag) & (mag) & & (mag) & (mag) \\
\hline
\textit{gri} & fitted & fitted & 0.162 $\pm$ 0.043 & 2.62 $\pm$ 0.19 & 0.045 $\pm$ 0.049 & 40 & 0.153 $\pm$ 0.031 & 0.155 $\pm$ 0.018 \\
\textit{co} & fitted & fitted & 0.140 $\pm$ 0.033 & 2.35 $\pm$ 0.19 & 0.044 $\pm$ 0.049 & 47 & 0.143 $\pm$ 0.032 & 0.168 $\pm$ 0.023 \\
\textit{coJH} & fitted & fitted & 0.069 $\pm$ 0.025 & 2.84 $\pm$ 0.17 & 0.061 $\pm$ 0.037 & 47 & 0.115 $\pm$ 0.020 & 0.131 $\pm$ 0.016 \\
\textit{coYJH} & fitted & fitted & 0.073 $\pm$ 0.025 & 2.75 $\pm$ 0.16 & 0.063 $\pm$ 0.036 & 47 & 0.121 $\pm$ 0.018 & 0.127 $\pm$ 0.016 \\
\textit{Y} & fixed & fixed & 0.000 & 0.00 & 0.052 $\pm$ 0.036 & 47 & 0.127 $\pm$ 0.024 & 0.126 $\pm$ 0.013 \\
\textit{J} & fixed & fixed & 0.000 & 0.00 & 0.040 $\pm$ 0.043 & 47 & 0.132 $\pm$ 0.023 & 0.148 $\pm$ 0.019 \\
\textit{H} & fixed & fixed & 0.000 & 0.00 & 0.036 $\pm$ 0.042 & 47 & 0.106 $\pm$ 0.030 & 0.146 $\pm$ 0.015 \\
\textit{YH} & fixed & fixed & 0.000 & 0.00 & 0.051 $\pm$ 0.037 & 47 & 0.105 $\pm$ 0.021 & 0.128 $\pm$ 0.014 \\
\textit{YJ} & fixed & fixed & 0.000 & 0.00 & 0.053 $\pm$ 0.036 & 47 & 0.116 $\pm$ 0.019 & 0.126 $\pm$ 0.016 \\
\textit{JH} & fixed & fixed & 0.000 & 0.00 & 0.039 $\pm$ 0.040 & 47 & 0.113 $\pm$ 0.023 & 0.141 $\pm$ 0.018 \\
\textit{YJH} & fixed & fixed & 0.000 & 0.00 & 0.050 $\pm$ 0.036 & 47 & 0.097 $\pm$ 0.021 & 0.127 $\pm$ 0.016 \\
\textit{co} & fitted & fixed & 0.130 $\pm$ 0.037 & 0.00 & 0.016 $\pm$ 0.062 & 47 & 0.213 $\pm$ 0.037 & 0.213 $\pm$ 0.025 \\
\textit{coJH} & fitted & fixed & 0.130 $\pm$ 0.027 & 0.00 & 0.039 $\pm$ 0.049 & 47 & 0.203 $\pm$ 0.028 & 0.169 $\pm$ 0.014 \\
\textit{coYJH} & fitted & fixed & 0.129 $\pm$ 0.022 & 0.00 & 0.053 $\pm$ 0.043 & 47 & 0.159 $\pm$ 0.027 & 0.149 $\pm$ 0.014 \\
\textit{Y} & fitted & fixed & 0.020 $\pm$ 0.000 & 0.00 & 0.044 $\pm$ 0.038 & 46 & 0.135 $\pm$ 0.028 & 0.130 $\pm$ 0.013 \\
\textit{YH} & fitted & fixed & 0.080 $\pm$ 0.030 & 0.00 & 0.063 $\pm$ 0.037 & 47 & 0.121 $\pm$ 0.023 & 0.131 $\pm$ 0.014 \\
\textit{YJ} & fitted & fixed & 0.020 $\pm$ 0.005 & 0.00 & 0.048 $\pm$ 0.037 & 47 & 0.114 $\pm$ 0.019 & 0.127 $\pm$ 0.016 \\
\textit{YJH} & fitted & fixed & 0.076 $\pm$ 0.037 & 0.00 & 0.065 $\pm$ 0.037 & 47 & 0.127 $\pm$ 0.024 & 0.131 $\pm$ 0.013 \\
\hline
\label{tab:long_Data}
\end{tabularx}
\end{table*}

\end{document}